# Fitting Cyclic Experimental Load-Deformation Data to The Pivot Hysteresis Model Using Genetic Algorithm

**MirSalar Kamari[1], Oğuz Güneş[2]**

1. Masters' Student, Department of Civil Engineering, Istanbul Technical University, kamari@itu.edu.tr
2. Assistant Professor, Department of Civil Engineering, Istanbul Technical University, ogunes@itu.edu.tr

**Abstract**

Understanding the linear or nonlinear relationship between load and deformation in structural materials or structural frames is a key to a proper and a well-represented simulation. This research is dedicated to model a cyclic load-deformation hysteresis relationship, captured from experimental results, and utilize it to represent the cyclic hysteresis data. The Genetic Algorithm is used to find the best parameters to introduce the model and to minimize the deviation between the simulation and the experimental results. In other words, the parameters associated with the loading response of any displacement pattern, are found, while minimizing the deviation between simulation loading response and the loading data carried out from the experiment. First, to reduce the data size recorded with measuring devices or Linear Variable Differential Transformers (LVDTs in short), a data resampling technic is proposed. Second, the resampled data is used in Genetic Algorithm to seek for the best parameters to describe the model. This method could be used to train models to predict the capacity and performance of a materials and frames when they are exposed to any deformation patterns. Numerous examples are shown at the end of the paper.

**Keywords:** Genetic Algorithm, Pivot hysteresis model, non-linear analysis, curve fitting, sap2000 API

## 1. Introduction

Genetic Algorithm (GA in short) is an optimization method based on heuristic search for the solutions that fit the best to the fitness criterion. GA optimization theory had been proposed by Joh Holland in the 1960s. Since then, it has been popular and widely been used in optimization problems and Its simple structure made it applicable in different scientific fields such as aerospace engineering [1], geophysics [2], material engineering [3], and financial marketing [4].

In civil structural engineering great number of researches has been carried out to obtain the optimal design of two dimensional frames using Genetic Algorithm form which Lee and Ahn [5] and Camp et al [6] are the most popular ones. Optimal design of RC frames was carried out by Rajeev and Krishnamoorthy [7] using a Simple Genetic Algorithm. Optimum design of a continues beam using GA were introduced by Govindaraj and Ramasm [8, 9].

In this particular research GA is used to find the parameters of a hysteresis load deformation model by minimizing the deviation between the experimental cyclic load-deformation results and that of the simulated model. First, the cyclic hysteresis recorded results are resampled to reduce the enormous size of the recorded data, and to remove the steady and repetitive data record. Second, backbone envelope load-deformation curve is extracted from the cyclic experimental results. Third, the arbitrary parameters are used to define the hysteresis model of the experimental cyclic results. These parameters are then optimized through GA, to minimize the deviation between the simulated and experimental results. In the following sections the process is explained in explicit details.

## 2. Cyclic data acquisition

The values of load and displacement history of specimen are recorded and measured during the experiment, thanks to Linear Variable Differential Transformers (LVDTs) devices. The raw data recorded from the LVDTs during the experiment period might have a very long dimension. This is because during the experiment period, the LVDTs are always measuring and recording the data, even though the change rate in the input parameters such as applying displacements to the experiment, are not tangible. For instance, a very large data could be recorded, when examinee crew tries to record the load versus displacement of a concrete frame due to excessive lateral excitation. This takes place at applying reversal loading, when the examinee crew will stop loading the frame, preparing for a reversal loading, while the LVDT instruments will keep storing steady load-deformation measures for the specimen. Analyzing and studying such a huge and abundant size of data is a very time-consuming and an exhausting task. Therefore, resampling of the data is a necessity, and in section four the resampling technics are introduced to prepare data record history to be used in GA. In the following section extracting backbone skeleton curve from cyclic load-deformation data is explored.

## 3. Obtaining idealized load-deformation backbone curve

Once that the experimental load-deformation cyclic data is recorded from any specimen, the backbone load-deformation relationship could be identified for it. The backbone curve represents the specimen inelastic non-linearity pattern for an excessive load or displacement setting. This loading pattern depicts the yielding and ultimate strength as well as the toughness of the specimen. Conventionally the load-deformation backbone relationship was extracted manually form the data, though, in this particular research this curve is extracted automatically from the experimental results.

To obtain the load-deformation relationship automatically from the experimental results the following steps are considered. First, the local maximums and minimums of load values for each cycle are extracted, and, each cycle is separated to be studied explicitly. Second, for each cycle, the maximum load value and its corresponding deformation value associated with each cycle is retrieved. After retrieving all load maximums of hysteresis data, to finalize calculation of the backbone curve, maximums points of each cycle are then sorted from their lowest to highest deformation value with an ascending rate. The following algorithm is developed to obtain the load deformation curve automatically, from the experimental load-deformation cyclic results.

| |
|---|
| **Input:** Displacement LVDT recorded array: *(Displacement)* |
|         Load LVDT recorded array: *(Load)* |
| **Output:** Backbone envelope curve's Displacement array: *(EnvelopeDisplacement)* |
|          Backbone envelope curve's Load array: *(EnvelopeLoad)* |
| (Obtaining the direction change in base-shear data record as the following) |
| **1 for** each consecutive member of (*Load$_i$*, *Load$_{i+1}$*) |
| **2**    $pp \leftarrow Load_i \times Load_{i+1}$ |
| **3 if** $pp < 0$ |
| **4**    **Append** *i* index of the *Load* array to *DirectionChenge* array |
| **5**  **end if** |
| **6 end for** |
| (Obtaining all of the local minima and maxima for *Load* array) |
| **7 for** each consecutive member of (*DirectionChenge$_i$*, *DirectionChenge$_{i+1}$*) |
| **8**   *Subset* ← cut *DirectionChenge$_i$* to *DirectionChenge$_{i+1}$* members of *Load* array |
| **9**   Set average of all *Subset* array members to *SubsetAve* |
| **10**  **if** *SubsetAve* > 0 |
| **11**    **Append** maximum of *Subset* array to *LoadMaxima* |
| **12**    *IndexVal* ← Return index of *LoadMaxima in Load* array |
| **13**    **Append** *IndexVal* to *LoadMaximaIndex* |
| **14**  **else** |
| **15**    **Append** minimum of *Subset* array to *LoadMinima* |
| **16**    *IndexVal* ← Return index of *LoadMinima in Load* array |
| **17**    **Append** *IndexVal* to *LoadMinimaIndex* |
| **18**  **end if** |
| **19 end for** |
| (Obtaining all corresponding maxima and minima for *Displacement* array) |
| **20 for** each member of (*LoadMaximaIndex$_i$*) |
| **21**   **Append** *LoadMaximaIndex'th* index of *Displacement* array to *DisplacementMaxima* |
| **22 end for** |
| **23 for** each member of (*LoadMinimaIndex$_i$*) |
| **24**   **Append** *LoadMinimaIndex'th* index of *Displacement* array to *DisplacementMinima* |
| **25 end for** |
| (Sort the resulting arrays) |
| **26 Append** *DisplacementMaxima* and *DisplacementMinima* to *EnvelopeDisplacement* |
| **27 Append** *LoadMaxima* and *LoadMinima* to *EnvelopeLoad* |
| **28 Sort** *EnvelopeDisplacement* and their corresponding *EnvelopeLoad* values with ascending rate |
| **29 Return** *EnvelopeDisplacement* and *EnvelopeLoad* |

**Figure 1: Algorithm for obtaining backbone curve from cyclic data**

The backbone curve is then idealized with seven points (three points at either positive and negative side, as well as one origin point). These three points at the positive and negative side of deformation axis are obtained as the following steps. First, to separate the elastic portion on the backbone curve on positive side, the origin is connected to lowest level of loading value, whose value exceeds the 65 percent of maximum loading value of the backbone curve in the positive deformation side. This will form the elastic portion of backbone, with two points. Second, on the positive side, the point whose loading value on the backbone curve is the maximum is taken as the third idealization point, and, finally, the point whose deformation is the maximum is taken as the fourth point on the idealization curve. The similar rules are applied to carry out the idealize backbone at the negative deformation side.

This idealization methodology has retrieved promising results, while carrying out the idealization backbone curve. This method works with load-deformation backbone data whose loading capacity is degraded through higher deformation levels. This behavior is mostly observed in concrete and masonry specimens, concrete frames with or without infill masonry walls while they are exposed to in-plane excessive lateral excitations.

The following Figure represents the structure of the algorithm to idealize the backbone curve.

**Input:** Backbone envelope curve's Displacement array: *(EnvelopeDisplacement)*
  Backbone envelope curve's Load array: *(EnvelopeLoad)*
**Output:** Idealized backbone envelope curve's displacement array: *(IdzEnvelopeDisplacement)*
  Idealized backbone envelope curve's Load array: *(IdzEnvelopeLoad)*

1 *IdzEnvelopeDisplacement$_{7\times1}$* ← zero vector with size of 7×1
2 *IdzEnvelopeLoad$_{7\times1}$* ← zero vector with size of 7×1
(returning the ultimate displacement points of backbone curve)
3 *IdzEnvelopeDisplacement $_7$* ← maximum(*EnvelopeDisplacement*)
4 *IdzEnvelopeLoad $_7$* ← corresponding *EnvelopeLoad* of maximum *EnvelopeDisplacement*
5 *IdzEnvelopeDisplacement $_1$* ← minimum(*EnvelopeDisplacement*)
6 *IdzEnvelopeLoad $_1$* ← corresponding *EnvelopeLoad* of maximum *EnvelopeDisplacement*
(returning the ultimate load points of backbone curve*)*
7 *IdzEnvelopeLoad $_6$* ← maximum(*EnvelopeLoad*)
8 *IdzEnvelopeDisplacement $_6$* ← corresponding *EnvelopeDisplacement* of maximum *EnvelopeLoad*
9 *IdzEnvelopeLoad $_2$* ← minimum(*EnvelopeLoad*)
10 *IdzEnvelopeDisplacement $_2$* ← corresponding *EnvelopeDisplacement* of maximum *EnvelopeLoad*
(returning the yield points of backbone curve*)*
11 **for** all ascending positive consecutive member of (*EnvelopeLoad$_i$*)
12  **if** *EnvelopeLoad$_i$* > 0.65 × maximum(*EnvelopeLoad*)
13   *IdzEnvelopeLoad $_5$* ← *EnvelopeLoad$_i$*
14   *IdzEnvelopeDisplacement $_5$* ← corresponding *EnvelopeDisplacement* of *EnvelopeLoad$_i$*
15  **end if**
16  **break**
17 **end for**
18 **for** all descending negative consecutive member of (*EnvelopeLoad$_i$*)
19  **if** *EnvelopeLoad$_i$* < 0.65 × minimum(*EnvelopeLoad*)
20   *IdzEnvelopeLoad $_3$* ← *EnvelopeLoad$_i$*
21   *IdzEnvelopeDisplacement $_3$* ← corresponding *EnvelopeDisplacement* of *EnvelopeLoad$_i$*
22  **end if**
20  **break**
21 **end for**
22 **Return** *IdzEnvelopeLoad* and *IdzEnvelopeDisplacement*

**Figure2: Algorithm to carry out the idealization of the backbone curve**

The backbone curve for two set of cyclic load-deformation data has been calculated automatically in the figure below, using the proposed method.

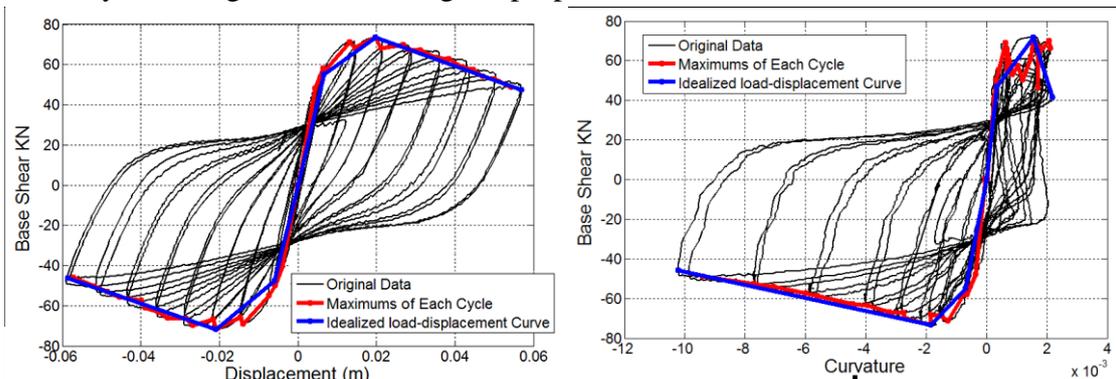

**Figure 3: Original LVDT load-deformation record versus calculated backbone curve**

Cyclic data in left hand side is experimental base-shear versus displacement result of a reinforced concrete bare frame, which was exposed to a displacement controlled in-plane lateral excitation, and the data in the right, is base-shear versus curvature cyclic result of one of the same frame's plastic joints, recorded during experiment.

## 4. Resampling strategies

The massive LVDT data record is resampled using two main strategies. First, a regular resampling, which restores indices of the load-deformation record with a regular pattern. Second, the elimination of steady record of the data is carried out.

To conduct a regular resampling, say total 10 values are about to be resampled with the reduction size of 2. Every index with step size of two are restored as the result of resampling process, and, 5 values are saved in resulting array. Figure below shows regular resampling technic.

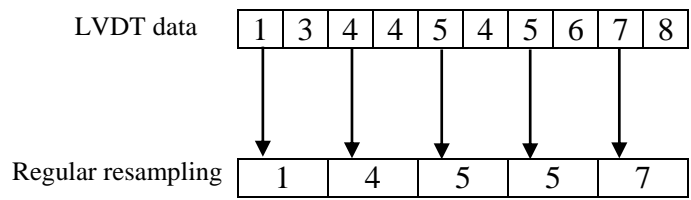

**Figure 4: Regular resampling the data with reduction (step) size of two**

The structure of the algorithm to carry out regular resampling can be shown as the figure below:

**Input:** Displacement LVDT record array with size of $n \times 1$: (*Dis*)
　　　　Load LVDT record array with size of $n \times 1$: (*Load*)
　　　　Reduction size: (*m*)
**Output:** Reduced size of displacement array (*RedDis*)
　　　　　Reduced size of load array (*RedLoad*)

1 *index* ←*1* to *n* with step size of *m*
2 **for** each member of *index* array, *j*
3 　**Append** *j*'th index of *Dis* array to *RedDis*
4 **end for**
5 **for** each member of *index* array, *j*
6 　**Append** *j*'th index of *Load* array to *RedLoad*
7 **end for**
8 **Return** *RedDis* and *RedLoad*

**Figure 5: The algorithm for regular data reduction**

While recording the load and deformation data with LVDT devices, due to irregular and non-steady nature of excitation application on specimen, the measuring devices will always record a fairly large size of data with non-steady change rate of input excitation. To reduce the size of data, and, to stabilize excitation change rate, the irregular resampling will have to be conducted. The following figure explores irregular resampling on a reduced data.

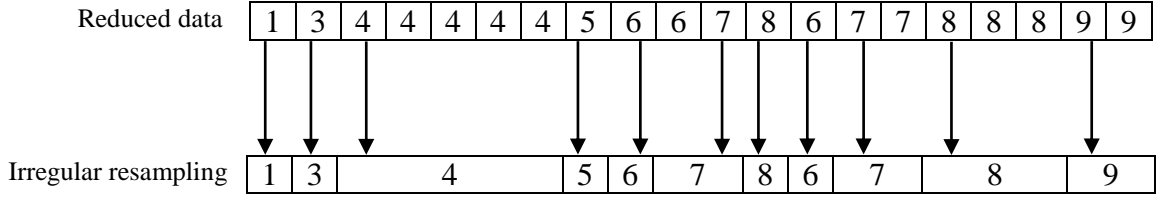

**Figure 6: An example of irregular resampling**

The structure of the algorithm to carry out the irregular resampling is shown as the following.

**Input:** Regularly reduced displacement record array with size of $n1 \times 1$: (*RedDis*)
Regularly reduced loading array record with size of $n1 \times 1$: (*RedLoad*)
Reduction size which is always greater than ten: *ResamplingScale*
Indices of load direction change: *ChangeDir*
**Output:** Resampled load array: *LoadINT,* and resampled displacement array: *DispINT*

(Initialize the parameters)
**1 Append** the last value of *RedDis* to *ChangeDir* matrix
**2** *LoadINT* ← [ ]
**3** *DispINT* ← [ ]
**4** *FirstDispVal* ← **round**(*RedDis$_{(1)}$*)
**5** *FirstDispIndex* ← 1
(Algorithm main body)
**6** *RedDis* ← *RedDis* × *ResamplingScale*
**7** *RedLoad* ← *RedLoad* × *ResamplingScale*
**8 round** all members of toward negative infinity
**9 for** the second to the last member of *changeDir $_i$*
**10  if** *FirstDispVal* < *ChangeDir$_{(i)}$*'th index of *RedDis*
**11**   *LastDispIndex* ← *ChangeDir$_{(i)}$*
**12**   *LastDispVal* ← *RedDis$_{(LastDispIndex)}$*
**13**   *Disploop* ← create a set of integers from *(FirstDispVal+1)* to *(LastDispVal)* Value with ascending rate
**14**   *DispFNC* ← create a set containing values of *RedDis$_{(FirstDispIndex)}$* to *RedDis$_{(LastCurvIndex)}$*
**15**   *DispFNC,Indices* ← **get** all the unique members of *DispFNC* and their indices (of first uniqueness)
**16**   *DispFNC* ← **get** members of *RedDis* with following indices: [*Indices+FirstDispIndex-1*]
**17**   *LoadFNC* ← **get** members of *RedLoad* with following indices: [*Indices+FirstDispIndex-1*]
**18**   *LoadLoop* ← **get** linear interpolation of *DispFNC* over *LoadFNC* with query points of *Disploop*
**19**   *FirstDispIndex* ← *LastDispIndex*
**20**   *FirstDispVal* ← *LastDispVal*
**21**   **Append** *Disploop* to *DispINT*
**22**   **Append** *LoadLoop* to *LoadINT*
**23  else if** *FirstDispVal* > *ChangeDir$_{(i)}$*'th index of *RedDis*
**24**   *LastDispIndex* ← *ChangeDir$_{(i)}$*
**25**   *LastDispVal* ← *RedDis$_{(LastDispIndex)}$*
**26**   *Disploop* ← create a set of integers from *(FirstDispVal-1)* to *(LastDispVal)* Value with descending rate
**27**   *DispFNC* ← create a set containing values of *RedDis$_{(FirstDispIndex)}$* to *RedDis$_{(LastCurvIndex)}$*
**28**   *DispFNC,Indices* ← **get** all the unique members of *DispFNC* and their indices (of first uniqueness)
**29**   *DispFNC* ← **get** members of *RedDis* with following indices: [*Indices+FirstDispIndex-1*]
**30**   *LoadFNC* ← **get** members of *RedLoad* with following indices: [*Indices+FirstDispIndex-1*]
**31**   *LoadLoop* ← **get** linear interpolation of *DispFNC* over *LoadFNC* with query points of *Disploop*
**32**   *FirstDispIndex* ← *LastDispIndex*
**33**   *FirstDispVal* ← *LastDispVal*
**34**   **Append** *Disploop* to *DispINT*
**35**   **Append** *LoadLoop* to *LoadINT*
**36  end if**
**37 end for**
**38** *LoadINT* ← *LoadINT* / *ResamplingScale*
**39** *DispINT* ← *DispINT* / *ResamplingScale*
**40 Return** *DispINT* and *LoadINT*

**Figure 7: Algorithm for irregular resampling**

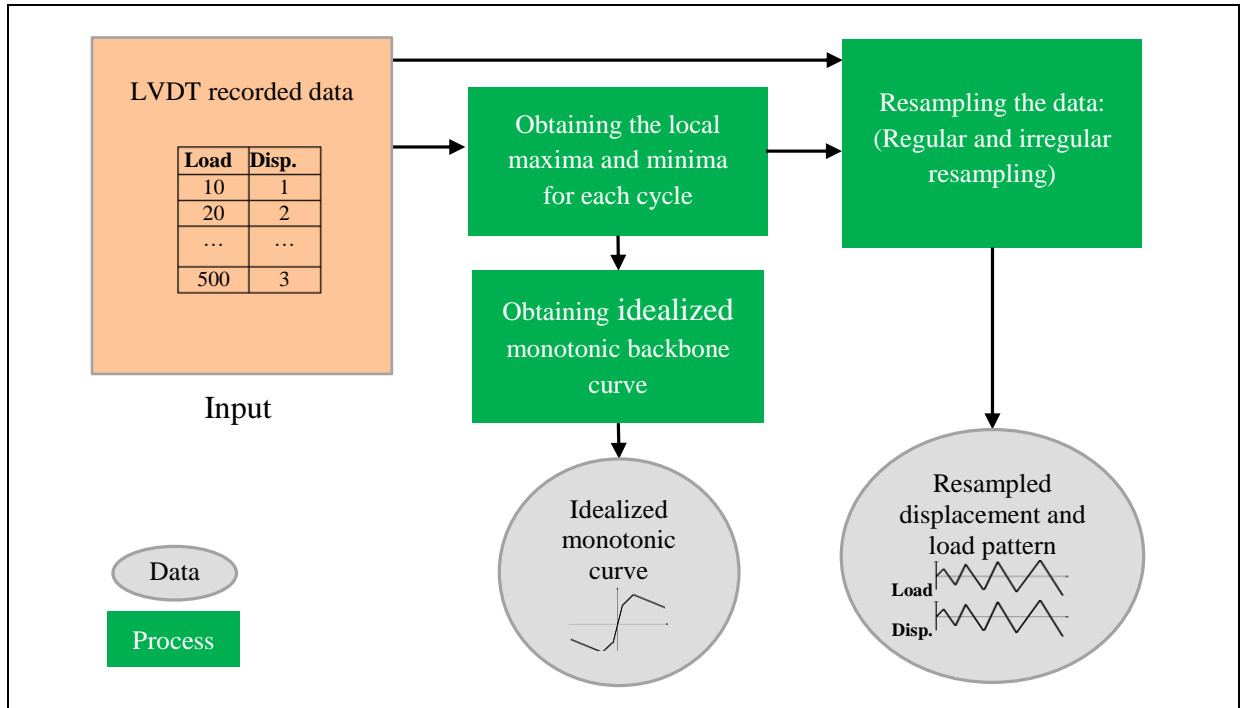

**Figure 8: Overview of algorithm of resamling and obtaing backbone curve of LVDT data**

## 5. Genetic Algorithm

Genetic Algorithm (GA in short) is a subcategory of evolutionary algorithms which are inspired by natural evolutions. At each iteration, each candidate solution is evaluated though fitness function which assesses the candidate's deviation with the demanding criterion. The best solution will be substitute with a candidate solution only if it has a better fit to the fitness function.

GA consists of individuals, who are the possible candidates at each step of iteration, and generations which are the populations that are evolving toward the optimal answer. GA basically uses each successful individual, who rated better in their generation though fitness function, as a parent to produce the population for the next generation. To converge to the optimal solution more quickly, GA produces the next generation based on the mutation and crossover function, which leads to a better diversity in population in the search for the most optimal answer.

After stabilizing change rates in experimental results data though resampling, this data is ready to be compared against with a computer generated data. Equal step of increase between displacement values, make it possible to generate the same deformation pattern with the same increment. For generated deformation pattern a loading pattern could be calculated through hysteresis rules, and this generated loading pattern could be compared against with that of experimental results to assess their deviation.

Pivot model containing five parameters is used to define hysteresis rules to generate the loading pattern form the deformation pattern. These five parameters are optimized thanks to GA to minimize the deviation between the generated load pattern and experimented load pattern. The optimization needs resampled load and deformation results as well as the skeleton backbone curve to deliver the Pivot hysteresis model's parameters that well define experimental results.

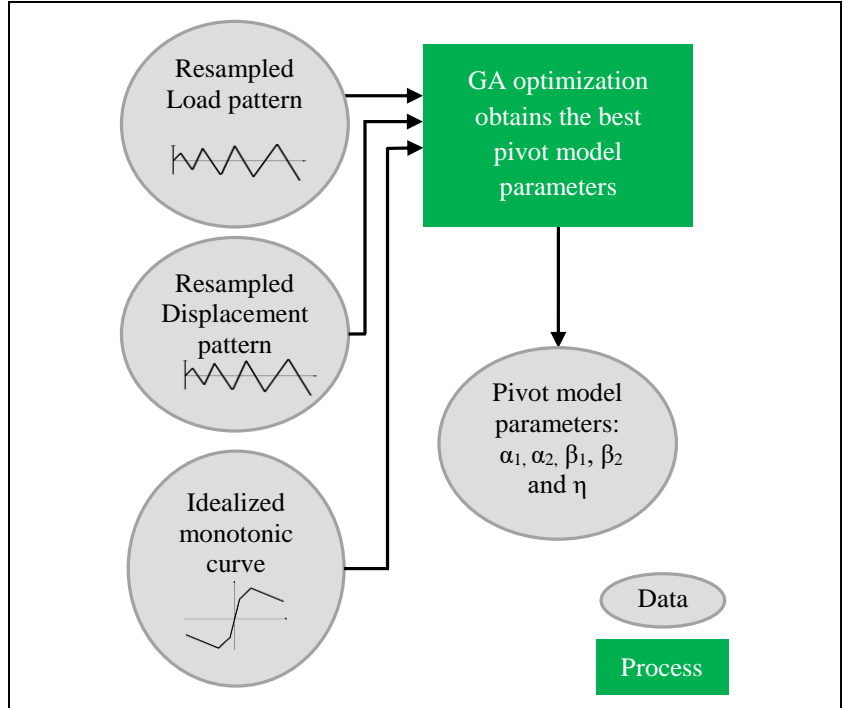

**Figure 9:** An overview of obtaining the best parameters of Pivot model while fitting the simulation to experimental results using Genetic Algorithm

## 6. Fitness functions

Given an idealized load-deformation curve and a displacement cyclic record, a load hysteresis response can be obtained using rules of Pivot model. The carried out hysteresis load response can be compared against the experimental load results to minimize deviation. It is worth mentioning that the dimension of the load-deformation of experimental results will have to be the same with that of simulated results. The following equation is used to assess the deviation between load response of the model and the experimental results;

$$\text{Deviation Score} = \sum_{i=1}^{n}(LoadResp_i - LoadExp_i)^2$$

where $LoadResp$ is load response record of the simulated model calculated based on its hysteresis rules with size of $n \times 1$, and $LoadResp_i$, is the $i$th member in $LoadResp$ array. $LoadExp$, is an array containing resampled experimental loading response with size of $n \times 1$, and $LoadExp_i$, is the $i$th member in $LoadExp$ array.

## 7. The use of Application Programming Interface (API) to integrate the software packages and to automate the parametric studies

Without a need to program a large number of codes to study and analyze the data corresponding to a given input, nowadays already written software packages can act like an engine under the hood, allowing the user to integrate them with a programming platform to request the inputs, and yield the outputs, through the Application Programming Interface (APIs). In other words, the sophisticated functionality of software packages can be manipulated thanks to APIs to study large number of inputs while doing iterations, optimizations, and parametric studies. This empowers researches, programmers and software developers to take advantage of the-state-of-the-art technologies, and implement their codes upon them to introduce a better software platform. With the power of the API in mind, it is no wonder that the Apple founder, Steve Jobs, builds his 3D map software upon Google Maps, though an API, to deliver a better software experience.

In this particular research Sap2000 has been integrated with MATLAB through an API to perform optimizations. None of the hysteresis rules were defined or coded but it was requested from Sap2000 from MATLAB to study different model's load responses of a given displacement, and Pivot rules parameters. Similar studies have been carried out by using Sap200 API integration [10], [11], [12], [13], [14]. Figure below shows the schematic structure of MATLAB and Sap2000 data request and retrieve.

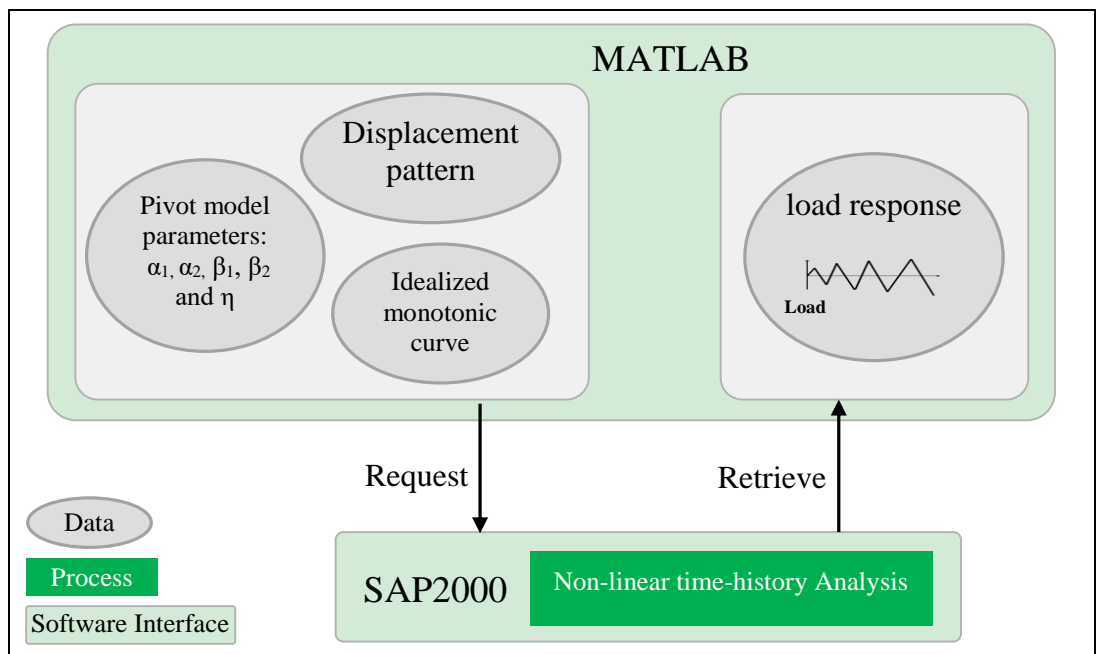

**Figure 10: Schematic structure of MATLAB and Sap2000 API**

## 8. Examples

In this section two examples are provided to fit the experimental data over proposed simulation method. First cyclic data represents the experimental results of a reinforced concrete bare frame, loaded with a lateral in-plane cyclic displacement pattern. The second example belong to cyclic experimental results of the same frame, though this time, curvature of one of its plastic joints have been plotted against base shear. Figures 11 (a), and 12 (a), show the resampled data versus the original data. The backbone curve and the idealized curve are also shown on these plots. Figures 11 (b), and 12 (b), show the resampled data versus simulated results coming from the GA processes. And Finally, Figures 11 (c), and 12 (c), show the GA convergence rate though generations.

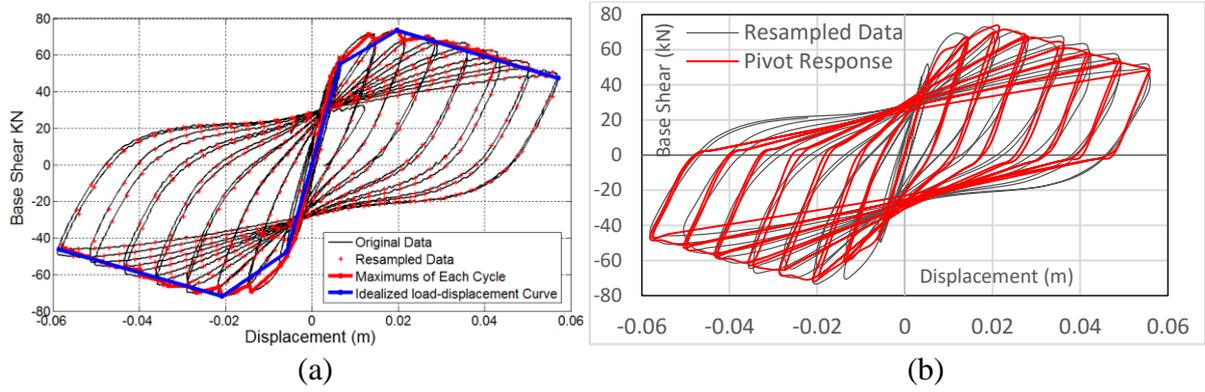

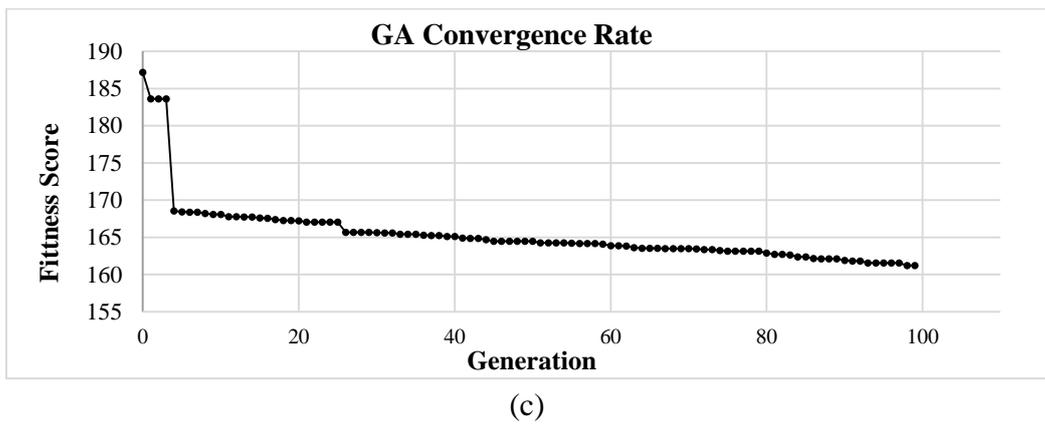

(c)

Figure 11: First example; (a) shows the original LVDT load-deformation data versus resampled data, (b) shows the resampled data versus the optimized pivot model through GA, and (c), represents GA convergence.

The calculation time was about an hour to optimize Pivot model to 500 resampled points. For this example, Pivot hysteresis model parameters have been found as:
$\alpha_1 = 19.82$, $\alpha_2 = 16.11$, $\beta_1 = 0.73$, $\beta_2 = 0.80$, and $\eta = 147.4$

Figure 12, shows the application of the proposed method over the second dataset. It represents resampled versus recorded load-deformation data, optimized Pivot model versus resampled data, and GA convergence rate through generations of optimization. The computational time was about an hour and a half. The resampled dataset was 637 set of points.

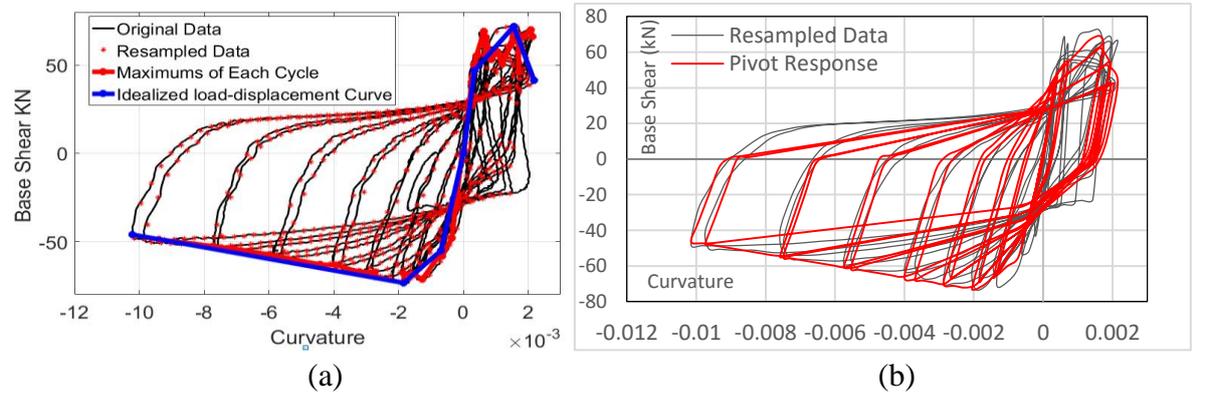

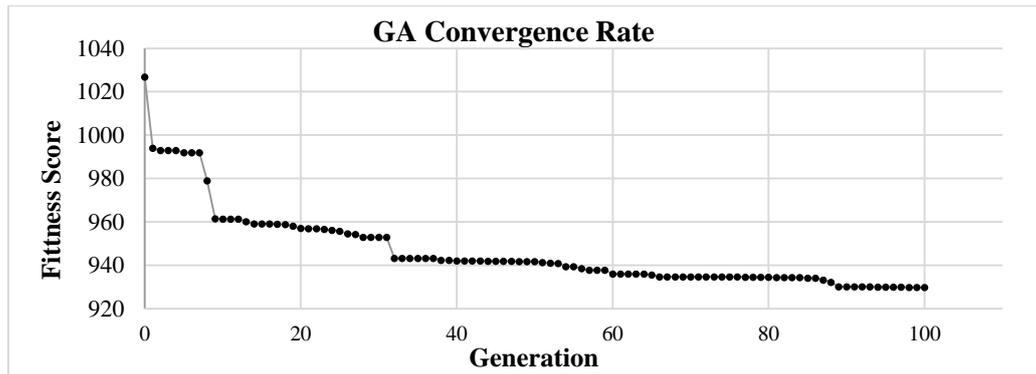

(c)

**Figure 12: Second example; (a) shows the original LVDT load-deformation data versus resampled data, (b) shows the resampled data versus the optimized pivot model through GA, and (c), represents GA convergence.**

For this example, Pivot hysteresis model parameters have been found as:
$\alpha_1 = 19.28$, $\alpha_2 = 25.47$, $\beta_1 = 1$, $\beta_2 = 0.73$, and $\eta = 447$

## 9. Acknowledgement

This work was partially supported by "Scientific and Technological Research Council of Turkey" (Turkish: Türkiye Bilimsel ve Teknolojik Araştırma Kurumu, TÜBİTAK) under contract No. 113M557. The data used in this research was collected at Middle East Technical University (METU), under supervision of Professor Polat Gülkan, Professor Haluk Sucuoğlu, and, Professor Barış Binici. The authors thank, and appreciate the effort of all the people at METU, who helped through data acquisition.